\documentclass[runningheads]{llncs}
\usepackage{graphicx}
\usepackage{xcolor}
\usepackage{amsmath,amssymb}
\usepackage{tabularx}
\usepackage{siunitx}
\DeclareMathOperator{\E}{\mathbb{E}}
\DeclareMathOperator{\KL}{\text{KL}}
\DeclareMathOperator{\CE}{\text{CE}}

\newcolumntype{Y}{>{\centering\arraybackslash}X}
\def\*#1{\mathbf{#1}}

\begin{document}
\title{PHiSeg: Capturing Uncertainty in Medical Image Segmentation}

\author{Christian F. Baumgartner\inst{1} \and Kerem C. Tezcan\inst{1} \and
Krishna Chaitanya\inst{1} \and Andreas M. H\"otker\inst{2} \and Urs J. Muehlematter\inst{2} \and Khoschy Schawkat\inst{2,4} \and Anton S. Becker\inst{2,3} \and Olivio Donati\inst{2} \and
Ender Konukoglu\inst{1}}
\authorrunning{C.F. Baumgartner et al}

\institute{Computer Vision Lab, ETH Z\"urich, Switzerland \and University Hospital Z\"urich, Switzerland \and Memorial Sloan Kettering Cancer Center, New York, USA \and Beth Israel Deaconess Medical Center, Harvard Medical School, Boston, USA}

\maketitle              
\begin{abstract}
Segmentation of anatomical structures and pathologies is inherently ambiguous. For instance, structure borders may not be clearly visible or different experts may have different styles of annotating. The majority of current state-of-the-art methods do not account for such ambiguities but rather learn a single mapping from image to segmentation. In this work, we propose a novel method to model the conditional probability distribution of the segmentations given an input image. We derive a hierarchical probabilistic model, in which separate latent variables are responsible for modelling the segmentation at different resolutions. Inference in this model can be efficiently performed using the variational autoencoder framework. We show that our proposed method can be used to generate significantly more realistic and diverse segmentation samples compared to recent related work, both, when trained with annotations from a single or multiple annotators. The code for this paper is freely available at \url{https://github.com/baumgach/PHiSeg-code}.

\end{abstract}

\section{Introduction}\label{sec:introduction}

Semantic segmentation of anatomical structures and pathologies is a crucial step in clinical diagnosis and many downstream tasks. The majority of recent automated segmentation methods treat the problem as a one-to-one mapping from image to output mask (e.g. ~\cite{ronneberger2015u}). However, medical segmentation problems are often characterised by ambiguities and multiple hypotheses may be plausible~\cite{warfield2002validation}. This is in part due to inherent uncertainties such as poor contrast or other restrictions imposed by the image acquisition, but also due to variations in annotation ``styles'' between different experts. To account for such ambiguities it is crucial that prediction systems provide access to the full distribution of plausible outcomes without sacrificing accuracy. Predicting only the most likely hypothesis may lead to misdiagnosis and may negatively affect downstream tasks. 

Recent work proposed to account for the uncertainty in the learned model parameters using an approximate Bayesian inference over the network weights~\cite{kendall2015bayesian}. However, it was shown that this method may produce samples that vary pixel by pixel and thus may not capture complex correlation structures in the distribution of segmentations~\cite{kohl2018probabilistic}. A different line of work accounts for the possibility of different outcomes by training an ensemble of $M$ networks~\cite{lakshminarayanan2017simple} or by training a single network with $M$ heads \cite{rupprecht2017learning}. Both approaches, however, can only produce a fixed number of hypotheses. This problem is overcome by the conditional variational autoencoder (cVAE), an extension of \cite{kingma2013auto} for modelling conditional segmentation masks given an input image~\cite{sohn2015learning}. Finally, the recently proposed probabilistic U-NET combines the cVAE framework with a U-NET architecture~\cite{kohl2018probabilistic}. The authors showed that, given ground-truth annotations from multiple experts, the method can produce an unlimited number of realistic segmentation samples. Moreover, the method was shown to outperform various related methods including network ensembles, $M$-heads~\cite{rupprecht2017learning} and the Bayesian SegNet~\cite{kendall2015bayesian}. 

\begin{figure}[t]\label{fig:graphical_model}
\centering
\includegraphics[width=0.99\textwidth]{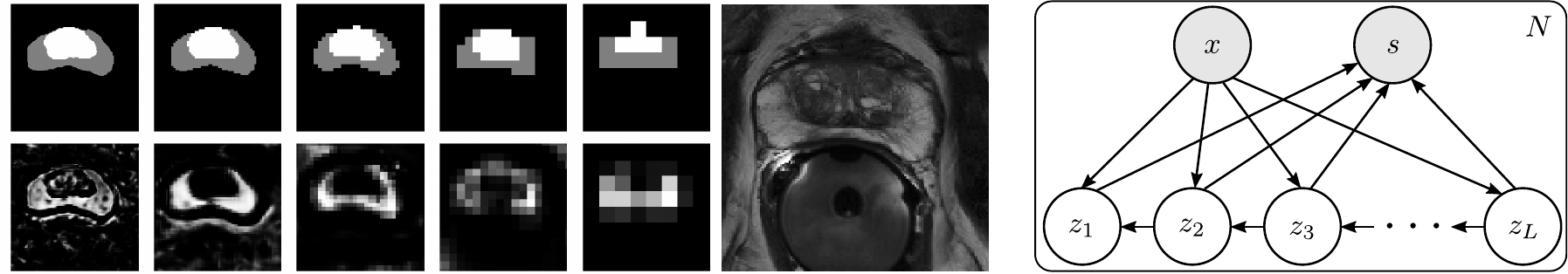}
\caption{(Left) Example of hierarchical segmentation generation with segmentation output at each level ($\hat{s}_\ell$) in the top row, right to left, and corresponding residual refinements to the prostate peripheral zone class in the bottom row. (Right) Corresponding graphical model (for $N$ independent samples).}
\end{figure}

However, as we will show, the probabilistic U-NET produces samples with limited diversity. We believe this may be due to the fact that stochasticity is only introduced in the highest resolution level of the U-NET, and because the network can choose to ignore the random draws from the latent space since it is only concatenated to the channels. In this work, we propose a novel hierarchical probabilistic model which can produce segmentation samples closely matching the ground-truth distribution of a number of annotators. Inspired by Laplacian Pyramids, the model generates image-conditional segmentation samples by generating the output at a low resolution and then continuously refining the distribution of segmentations at increasingly higher resolutions. In contrast to prior work, the variations on \emph{each resolution level} are governed by a separate latent variable, thereby avoiding the problems mentioned above. This process is illustrated in Fig. \ref{fig:graphical_model}. We show that compared to recent work, our proposed Probabilistic Hierarchical Segmentation (PHiSeg) produces samples of significantly better quality for two challenging segmentation tasks, both, when trained with multiple annotations, and a single annotation per image. Furthermore, the mean prediction of our model performs on par with the standard U-NET in terms of segmentation accuracy. 

\section{Methods}

We start by assuming that the segmentations $s$ given an input image $x$ are generated from $L$ levels of latent variables $z_\ell$ according to the graphical model shown in Fig. \ref{fig:graphical_model}. Thus, the conditional distribution $p(s|x)$ is given by the following expression for the general case of $L$ latent levels: 
\begin{equation}\label{eq:graphical_model}
p(\*s|\*x) = \int p(\*s|\*z_1, \dots, \*z_L)p(\*z_1|\*z_2,\*x)\cdots p(\*z_{L-1}|\*z_L,\*x)p(\*z_L|\*x)d\*z_1\cdots d\*z_L.
\end{equation}
We further assume that each latent variable $z_\ell$ is responsible for modelling the conditional target segmentations at $2^{-\ell+1}$ of the original image resolution (e.g. $\*z_1$ and $\*z_3$ model the segmentation at the original and at $1/4$ of the original resolution, respectively.). This does not result from the graphical model itself but is rather enforced by our implementation thereof as will become clear shortly. 

\begin{figure}[t]\label{fig:architecture} 
\centering
\includegraphics[width=0.99\textwidth]{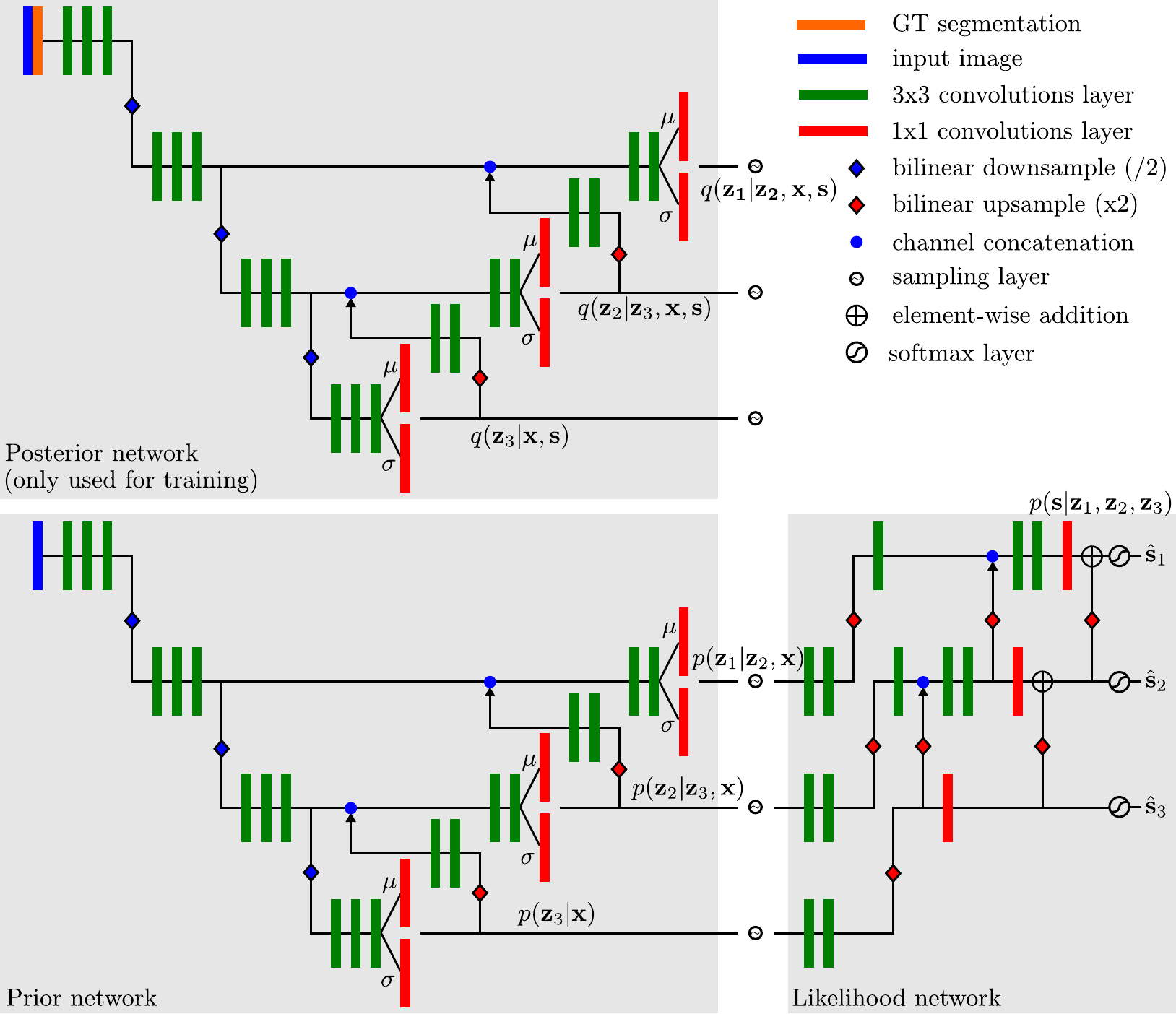}  
\caption{Schematic network architecture of the proposed method for $L=3$ latent levels and 4 resolution levels.}
\end{figure}

We aim to approximate the posterior distribution of $p(\*z|\*s,\*x)$ using a variational approximation $q(\*z|\*s,\*x)$ where we used $\*z$ to denote $\{\*z_1,\dots,\*z_L\}$. It can be shown that $\log p(\*s|\*x) = \mathcal{L}(\*s|\*x) + \KL(q(\*z|\*s,\*x)||p(\*z|\*s,\*x))$, where $\mathcal{L}$ denotes the \emph{evidence lower bound}, and $\KL(\cdot, \cdot)$ the Kullback-Leibler divergence~\cite{kingma2013auto,sohn2015learning,kohl2018probabilistic}. Since $\KL(\cdot, \cdot)\geq0$, $\mathcal{L}$ is a lower bound on the conditional log probability with equality when the approximation $q$ matches the posterior exactly. Using the decomposition in Eq.~\ref{eq:graphical_model} we find that for our model
\begin{equation}\label{eq:elbo}
\begin{split}
\mathcal{L} = & \E_{q(\*z_1,\dots,\*z_L|\*x,\*s)} \left[ \log p(\*s|\*z_1,\dots,\*z_L)\right] -\alpha_L  \KL\left[ q(\*z_L|\*s,\*x)||p(\*z_L|\*x)\right] \\
               - & \sum_{\ell=1}^{L-1} \alpha_\ell\E_{q(\*z_{\ell+1}|\*s,\*x)} \left[ \KL\left[ q(\*z_\ell|\*z_{\ell+1}, \*s, \*x) || p(\*z_\ell|\*z_{\ell+1}, \*x) \right] \right],
\end{split}
\end{equation}
with $\alpha_\ell=1$. A complete derivation can be found in Appendix~A. The $\alpha_\ell$ are additional heuristic variables which we introduced to help account for dimensionality differences between the $\*z_\ell$ (explained below). Following standard practice we parametrise the prior and posterior distributions as axis aligned normal distributions $\mathcal{N}(\*z|\mu, \sigma)$. Specifically, we define
\begin{align} 
p(\*z_\ell|\*z_{\ell+1}, \*x) &= \mathcal{N}\left(\*z|\phi^{(\mu)}_\ell(\*z_{\ell+1},\*x), \phi^{(\sigma)}_\ell(\*z_{\ell+1},\*x)\right)  \\  
q(\*z_\ell|\*z_{\ell+1}, \*x, \*s) &= \mathcal{N}\left(\*z|\theta^{(\mu)}_\ell(\*z_{\ell+1}, \*s, \*x), \theta^{(\sigma)}_\ell(\*z_{\ell+1}, \*s, \*x)\right),  
\end{align}
where the $\phi, \theta$ are functions parametrised by neural networks. Note that in contrast to the variational autoencoder~\cite{kingma2013auto}, the $p(\*z_\ell|\cdot, \*x)$ are also parametrised by neural networks similar to~\cite{kohl2018probabilistic,sohn2015learning}. Lastly, we model $p(\*s|\*z)$ as the usual categorical distribution with parameters (i.e. softmax probabilities) predicted by another neural network. By parametrising all distributions using neural networks, this can be seen as a hierarchical conditional variational autoencoder with the posteriors $q(\*z_\ell|\cdot, \*s, \*x)$ and priors $p(\*z_\ell|\cdot, \*s)$ encoding $\*x$ and $\*s$ into latent representations $\*z_\ell$, and the likelihood $p(\*s|\*z)$ acting as the decoder. Our implementation of this model using a neural network for $L=3$ is shown in Fig.~\ref{fig:architecture}. In that figure it can be seen that the total number of resolution levels of the network (i.e. number of downsampling steps plus one) can be larger than the number of latent levels. The example in Fig.~\ref{fig:architecture} has a total of 4 resolution levels, of which only $L=3$ are latent levels. We obtained the best results with 7 total resolution levels of which $L=5$ are latent levels. The prior and posterior nets have identical structure but do not share any weights. Similar to previous work, all three subnetworks are used for training but testing is performed by using only the prior and the likelihood networks~\cite{kingma2013auto,sohn2015learning,kohl2018probabilistic}. 

From Fig. \ref{fig:architecture} it can be seen that latent variables $\*z_\ell$ will form the skip connections in a U-NET-like architecture. However, unlike \cite{ronneberger2015u} and \cite{kohl2018probabilistic}, each skip connection corresponds to a latent variable $\*z_\ell$ such that no information can flow from the image to the segmentation output without passing a sampling step. We do not map the latent variables to a 1-D vector but rather choose to keep the structured relationship between the variables. We found that this substantially improves segmentation accuracy. As a result, latent variable $\*z_\ell$ has a dimensionality of $r_x 2^{-\ell+1}\times r_y 2^{-\ell+1} \times D$, where $D$ is a hyper-parameter and $D=2$ for all experiments, and $r_x, r_y$ are the dimensions of the input images. The latent variable $\*z_\ell$ is limited to modelling the data at $2^{-\ell+1}$ of the original resolution due to the downsampling operations before it. It then passes up the learned representation to the latent space embedding above ($\*z_{\ell-1}$) to perform a refinement at double the resolution. This continues until the top level is reached. To further enforce this behaviour the likelihood network is designed to generate only \emph{residual} changes of the segmentation masks for all $\*z_\ell$ except the bottom one. This is achieved through the addition layers before the outputs (see Fig.~\ref{fig:architecture}). Our model bears some resemblance to the Ladder Network~\cite{valpola2015neural} which is also a hierarchical latent variable model where inference results in an autoencoder with skip connections. Our work differs substantially from that work in how inference is performed. Furthermore, to our knowledge, the Ladder Network was never applied to structured prediction problems. 

{\bf Training and Predictions:} We aim to find the neural network parameters which maximise the lower bound $\mathcal{L}$ in Eq. \ref{eq:elbo}. The analytical form of the individual terms is prescribed by our model assumptions: since the posterior and prior both are modelled by normal distributions, the KL terms can be calculated analytically~\cite{kingma2013auto}. Our choice of likelihood results in a cross entropy term $\CE(\hat{\*s}_1, \*s_{gt})$, with $\hat{\*s}_1$ the predicted segmentation and $\*s_{gt}$ the corresponding ground-truth. Similar to previous work we found that it is sufficient to evaluate all of the expectations using a single sample~\cite{kingma2013auto}. Two deviations from the above theory were necessary for stable training. First, the magnitude of the KL terms depends on the dimensionality of $\*z_\ell$. However, since the dimensionality of $\*z_\ell$ in our model grows with $O(2^\ell)$, this led to optimisation problems. To counteract this, we heuristically set the weights $\alpha_\ell=2^{\ell-1}$ in Eq. \ref{eq:elbo}. Secondly, to enforce the desired behaviour that $\*z_\ell$ should only model the data at its corresponding resolution, we added deep supervision to the output of each resolution level ($\hat{\*s}_\ell$ in Fig. \ref{fig:architecture}). The cost function used for this is again the cross entropy loss, $\CE(ups(\hat{\*s}_\ell), \*s_{gt})$ for $\ell > 1$, where $ups(\cdot)$ denotes a nearest neighbour upsampling to match the size of $\*s_{gt}$. While $\*z_\ell$ can only model the data at a certain resolution, it may ignore this responsibility and focus only on matching the prior and posterior. Deep supervision effectively prevents this behaviour.

We trained the model using the Adam optimiser with a learning rate of $10^{-3}$ and a batch-size of 12. We used batch-normalisation on all non-output layers. All models were trained for 48 hours on a NVIDIA Titan Xp GPU and the model with the lowest total loss on a held-out validation set was selected. 

After the model is trained, segmentation samples for an input image $\*x$ can be generated by first obtaining samples $\*z_\ell$ using the prior network and then decoding them using the likelihood network. 

\section{Experiments and Results}  

We evaluated our method on two datasets: 1) the publicly available LIDC-IDRI dataset which comprises 1018 thoracic CT images with lesions annotated by 4 radiologists~\cite{armato2011lung}. Similar to \cite{kohl2018probabilistic} we extracted square 2D patches of size $128\times128$ pixels such that each patch was centred on a lesion. 2) We also evaluated our method on an in-house prostate MR dataset of 68 patients acquired with a transverse T2-weighted sequence (in-plane resolution $0.1875\times0.1875$\,\si{mm^2} and slice thickness $3.3$\,\si{mm}). The transition and peripheral zones were manually annotated by 4 radiologists and 2 non-radiologists. We processed the data slice-by-slice (approx. 25 slices per volume), where we resampled each slice to a resolution of $0.6\times0.6$\,\si{mm^2} and took a central crop of size $192\times192$. We divided both datasets into a training, testing and validation set using a random 60-20-20 split. 

For all experiments we compared our method (PHiSeg) with $L=5$ latent levels and a total of 7 resolution levels to the probabilistic U-NET~\cite{kohl2018probabilistic}. In order to exclude network capacity as an explanation for performance differences, we aimed to model our network components as closely as possible after the probabilistic U-NET. We used batch normalisation layers for both methods which deviates from \cite{kohl2018probabilistic} but did not affect the results negatively. Furthermore, to demonstrate that modelling the segmentation problem at multiple resolution levels is beneficial, we also compared against a variation of PHiSeg with only $L=1$ latent levels (i.e. no skip connections or latent space hierarchy). Lastly, for some experiments we compared to a deterministic U-NET using the same architecture as for the probabilistic U-NET but with no stochastic components. 

\begin{figure}[t]\label{fig:samples_prostate} 
\centering
\includegraphics[width=0.99\textwidth]{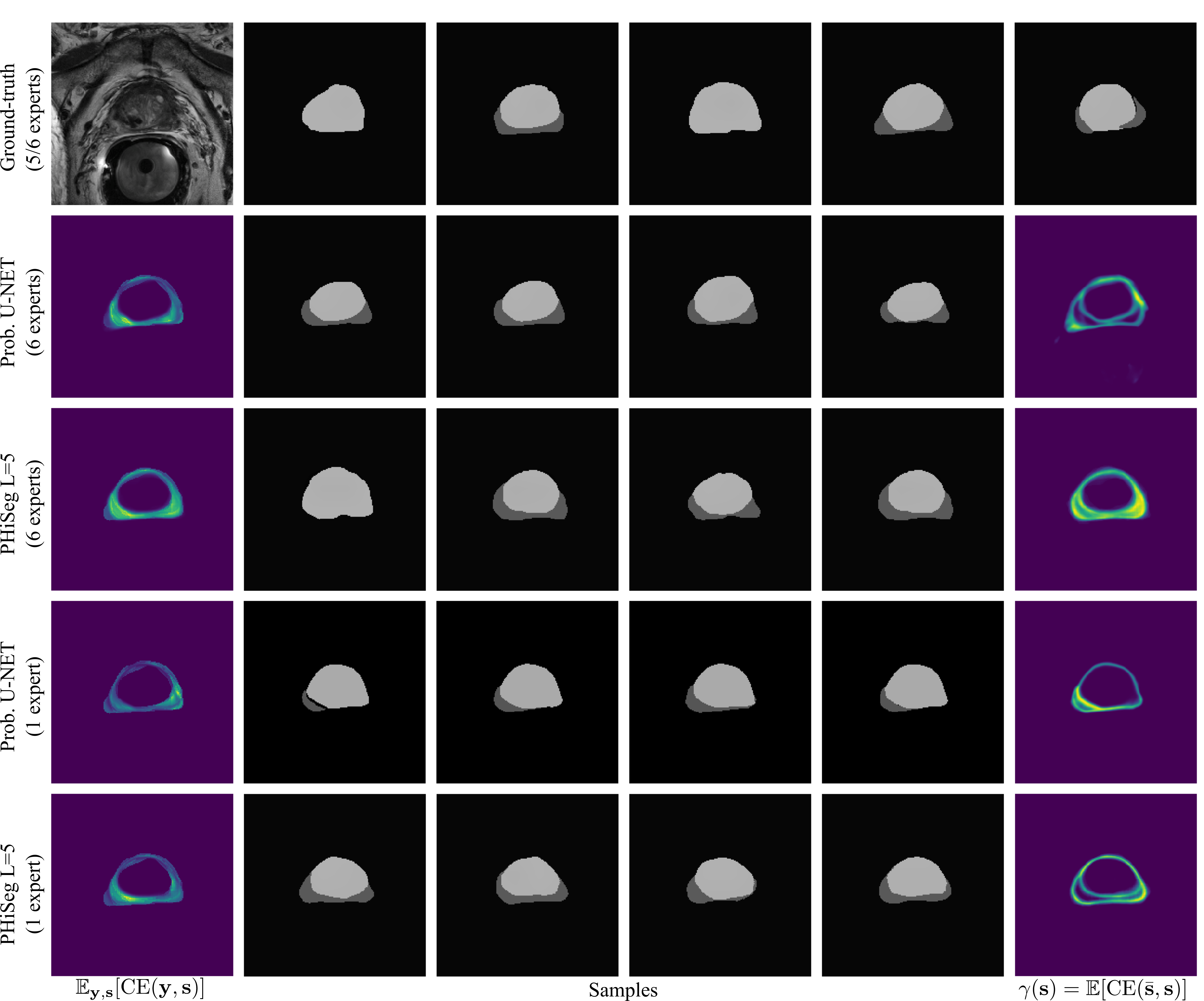}  
\caption{Ground-truth annotations and samples for two of the evaluated methods trained with masks from 6 or 1 experts(s). Average error maps $\E_{\*y,\*s}[\CE(\*y, \*s)]$ and $\gamma$-maps ($\E[\CE(\bar{\*s}, \*s)]$) for each model are shown in the left and right-most column, respectively.}
\end{figure}

We evaluated the techniques in two experiments. First, we trained the methods using the masks from all available annotators, where in each batch we randomly sampled one annotation per image. We were interested in assessing how closely the distribution of generated samples matched the distribution of ground-truth annotations. To this end, we used the generalised energy distance $ D^2_\text{GED}(p_{gt}, p_{\*s}) = 2\E[d(\*s,\*y)] -\E[d(\*s,\*s')] - \E[d(\*y,\*y')], $ where $d$ is 1 minus the intersection over union, i.e. $d(\cdot, \cdot) = 1-\text{IoU}(\cdot, \cdot)$, and $\*s,\*s',\*y,\*y'$ are samples from the learned distribution $p_\*s$, and ground-truth distribution $p_{gt}$~\cite{kohl2018probabilistic}. The GED reduces the sample quality to a single, easy-to-understand number but, as a consequence, cannot be interpreted visually. Therefore, we additionally aimed to produce pixel-wise maps showing variability among the segmentation samples. We found the expected cross entropy between the mean segmentation mask and the samples to be a good measure, i.e. $\gamma(s_i) = \E[\CE(\bar{s_i}, s_i)]$ with $i$ the pixel position and $\bar{s_i}$ the mean prediction. $\gamma$ is statistically similar to variance with the L2-distance replaced by CE. However, we believe it is more suitable for measuring segmentation variability. Examples of our $\gamma$-maps along with sample segmentations are shown in Fig. \ref{fig:samples_prostate}. We quantify how well the $\gamma$-maps for each method predict regions with large uncertainty using the average normalised cross correlation (NCC) between the $\gamma$-maps and the CE error maps obtained with respect to each annotator: 
\begin{equation}
\mathcal{S}_\text{NCC}(p_{gt}, p_{\*s}) = \E_{\*y\sim p_{gt}} \left[ \text{NCC}(\E_{\*s\sim p_{\*s}}[\CE(\bar{\*s},\*s)], \E_{\*s\sim p_{\*s}}[\CE(\*y,\*s)]) \right].
\end{equation} 
Results for both $D^2_\text{GED}$ and $ \mathcal{S}_\text{NCC}$ are shown in the top part of Tab.~\ref{tab:results}. All measures were evaluated with 100 samples drawn from the learned models. 

Secondly, we set out to investigate the models' ability to infer the inherent uncertainties in the annotations from \emph{just one} annotation per training image. To this end, we trained the above models by using only the annotations of a single expert. For the evaluation we then computed the $D^2_\text{GED}$ and $\mathcal{S}_\text{NCC}$ using all available annotators. Additionally, we evaluated the models in terms of conventional Dice score evaluated with masks from the single annotator as ground-truth. To get a single prediction from the probabilistic models we used $\bar{\*s}$. This allowed us to obtain an indication of conventional segmentation accuracy. The results are shown in the bottom part of Tab.~\ref{tab:results}. 
\begin{table*}[th!]
\caption{Quantitative results for all metric. Statistically significant improvements ($p<0.01$ with paired student's $t$-test) over all other methods are indicated in bold.}
\label{tab:results}
\centering
\begin{tabularx}{\textwidth}{l | c | YYY | YYY }
& &\multicolumn{3}{c|}{\textbf{LIDC-IDRI}}&\multicolumn{3}{c}{\textbf{Prostate dataset}} \\
 & \#\,experts & $ D^2_\text{GED}$ & $ \mathcal{S}_\text{NCC}$  & Dice & $ D^2_\text{GED}$ & $ \mathcal{S}_\text{NCC}$ & Dice\\
\hline
Prob. U-NET    & All & 0.2393       & 0.7749       & -      & 0.1322      & 0.7763      & -      \\
PHiSeg ($L=1$) & All & 0.2934       & 0.7944       & -      & 0.1608      & 0.7452      & -      \\
PHiSeg ($L=5$) & All & {\bf 0.2248} & {\bf0.8453}  & -      & {\bf0.0864} & {\bf0.8185} & -      \\
\hline
Det. U-NET     & 1   & -            & -            & 0.5297 & -           & -           & 0.8364 \\
Prob. U-NET    & 1   & 0.4452       & 0.5999       & 0.5238 & 0.2198      & 0.6022      & 0.8290 \\
PHiSeg ($L=1$) & 1   & 0.4695       & 0.6013       & 0.5275 & 0.2462      & 0.6683      & 0.7942 \\
PHiSeg ($L=5$) & 1   & {\bf 0.3225} & {\bf0.7337}  & 0.5408 & 0.2044      & {\bf0.6917} & 0.8540 \\
\hline
\end{tabularx}
\end{table*}

We observed that when using all annotators for training, PHiSeg ($L=5$) produced significantly better $ D^2_\text{GED}$ and $\mathcal{S}_\text{NCC}$ scores compared to all other methods. This can be observed qualitatively in Fig.~\ref{fig:samples_prostate} for a prostate slice with large inter-expert disagreements. Both, the prob. U-NET and PHiSeg ($L=5$) produced realistic samples but PHiSeg ($L=5$) was able to capture a wider variability. Furthermore, as indicated by the high $\mathcal{S}_\text{NCC}$ values, PHiSeg's ($L=5$) $\gamma$-maps were found to be very predictive of where in the image the method's average prediction errors will occur. Similar results were obtained when training with only one annotator. We noticed that in this scenario the prob. U-NET may in some cases fail to learn variation in the data and revert back to an almost entirely deterministic behaviour (see fourth row in Fig.~\ref{fig:samples_prostate}). We believe this can be explained by the prob. U-NET's architecture which, in contrast to our method, allows the encoder-decoder structure to bypass the stochasticity. While our method also predicted smaller variations in the samples, they were still markedly more diverse. The lower performance of PhiSeg ($L=1$) indicates that using multiple resolution levels is crucial for our method. More samples for the prostate and LIDC-IDRI datasets can be found in Appendix~B. From Tab.~\ref{tab:results} it can be seen that no significant differences between the Dice scores were found for any of the methods (except PHiSeg's ($L=1$)), including the det. U-NET. From this we conclude that neither PhiSeg ($L=5$) nor the prob. U-NET suffer in segmentation performance due to their stochastic elements. 

\section{Discussion and Conclusion}

We introduced a novel hierarchical probabilistic method for modelling the conditional distribution of segmentation masks given an input image. We have shown that our method substantially outperforms the state-of-the-art on a number of metrics. Furthermore, we demonstrated that PHiSeg was able to predict its own errors significantly better compared to previous work. We believe that proper modelling of uncertainty is indispensable for clinical acceptance of deep neural networks and that having access to the segmentation's probability distribution will have applications in numerous downstream tasks. 

\subsubsection*{Acknowledgements}
This work was partially supported by the Swiss Data Science Center. One of the Titan X Pascal used for this research was donated by the NVIDIA Corporation.

\bibliographystyle{splncs04}
\bibliography{references}

\begin{thebibliography}{10}
\providecommand{\url}[1]{\texttt{#1}}
\providecommand{\urlprefix}{URL }
\providecommand{\doi}[1]{https://doi.org/#1}

\bibitem{armato2011lung}
Armato, S.G., McLennan, G., Bidaut, L., McNitt-Gray, M.F., Meyer, C.R., Reeves,
  A.P., Zhao, B., Aberle, D.R., Henschke, C.I., Hoffman, E.A., et~al.: The lung
  image database consortium {(LIDC)} and image database resource initiative
  {(IDRI)}: a completed reference database of lung nodules on {CT} scans. Med.
  Phys.  \textbf{38}(2),  915--931 (2011)

\bibitem{kendall2015bayesian}
Kendall, A., Badrinarayanan, V., Cipolla, R.: Bayesian segnet: Model
  uncertainty in deep convolutional encoder-decoder architectures for scene
  understanding. arXiv:1511.02680  (2015)

\bibitem{kingma2013auto}
Kingma, D., Welling, M.: Auto-encoding variational bayes. arXiv:1312.6114
  (2013)

\bibitem{kohl2018probabilistic}
Kohl, S., Romera-Paredes, B., Meyer, C., De~Fauw, J., Ledsam, J.R., Maier-Hein,
  K., Eslami, S.A., Rezende, D.J., Ronneberger, O.: A probabilistic {U-Net} for
  segmentation of ambiguous images. In: Proc. NIPS. pp. 6965--6975 (2018)

\bibitem{lakshminarayanan2017simple}
Lakshminarayanan, B., Pritzel, A., Blundell, C.: Simple and scalable predictive
  uncertainty estimation using deep ensembles. In: Proc. NIPS. pp. 6402--6413
  (2017)

\bibitem{ronneberger2015u}
Ronneberger, O., Fischer, P., Brox, T.: U-net: Convolutional networks for
  biomedical image segmentation. In: Proc. MICCAI. pp. 234--241. Springer
  (2015)

\bibitem{rupprecht2017learning}
Rupprecht, C., Laina, I., DiPietro, R., Baust, M., Tombari, F., Navab, N.,
  Hager, G.D.: Learning in an uncertain world: Representing ambiguity through
  multiple hypotheses. In: Proc. CVPR. pp. 3591--3600 (2017)

\bibitem{sohn2015learning}
Sohn, K., Lee, H., Yan, X.: Learning structured output representation using
  deep conditional generative models. In: Proc. NIPS. pp. 3483--3491 (2015)

\bibitem{valpola2015neural}
Valpola, H.: From neural {PCA} to deep unsupervised learning. In: Persp Neural
  Comp, pp. 143--171. Elsevier (2015)

\bibitem{warfield2002validation}
Warfield, S.K., Zou, K.H., Wells, W.M.: Validation of image segmentation and
  expert quality with an expectation-maximization algorithm. In: Proc. MICCAI.
  pp. 298--306. Springer (2002)

\end{thebibliography}

\appendix
\renewcommand\thefigure{\thesection.\arabic{figure}}  

\section{Derivation of the Evidence Lower Bound (ELBO)}
\label{sec:appendix_proof}

In this section, we derive the ELBO given in Eq.~2 of the main article based on our model assumptions given in Fig.~1 (right) and Eq.~1 in the main article. We start from the well known decomposition for the variational lower bound for the conditional probability distribution $p(\*s|\*x)$\footnote{This can easily be derived by starting with $\KL(q(\*z|\*s,\*x)||p(\*z|\*s,\*x))$, replacing $p(\*z|\*s,\*x)=\tfrac{p(\*s|\*z,\*x)p(\*z|\*x)}{p(\*s|\*x)}$, and splitting the log term.}. Defining $\*z = \{\*z_1,\dots,\*z_L\}$ we can write:
\begin{equation}
\log p(\*s|\*x) = \mathcal{L}(\*s|\*x) + \KL(q(\*z|\*s,\*x)||p(\*z|\*s,\*x)) 
\end{equation}
with 
\begin{equation}\label{eq:var_decomp}
\mathcal{L}(\*s|\*x) = \E_{q(\*z|\*s,\*x)}\left[ p(s|\*z,\*x) \right] - \KL\left[ q(\*z|\*s,\*x) || p(\*z|\*x) \right] 
\end{equation}
Since $ \KL(q(\*z|\*s,\*x)||p(\*z|\*s,\*x)) $ is always positive $\mathcal{L}(\*s|\*x)$ is a lower bound on $\log p(\*s|\*x)$ with equality when $q(\*z|\*s,\*x)=p(\*z|\*s,\*x)$.

The right-most KL-divergence term in Eq. \ref{eq:var_decomp} can be further rewritten as 
\begin{equation}\label{eq:plugged_in_model}
\begin{split}
& \KL\left[ q(\*z_1,\dots,\*z_L|\*s,\*x)\,||\,p(\*z_1,\dots,\*z_L|\*x) \right] \\
& = \KL\left[q(\*z_L|\*s,\*x)q(\*z_1,\dots,\*z_{L-1} |\*z_{L},\*s,\*x) \,\middle\vert\middle\vert\, p(\*z_L|\*x)p(\*z_1,\dots,\*z_{L-1} |\*z_{L},\*x)\right] \\
& =\KL\left[ q(\*z_L|\*s,\*x)\,||\,p(\*z_L|\*x) \right] + \\
& ~~~\, \int \dots \int q(\*z_1,\dots,\*z_L|\*s,\*x) \log \frac{q(\*z_1,\dots,\*z_{L-1} |\*z_{L},\*s,\*x)}{p(\*z_1,\dots,\*z_{L-1} |\*z_{L},\*x)} d\*z_1 \dots d\*z_L,  \\
\end{split}
\end{equation}
where in the first equality we used the definition of our graphical model to separate out $\*z_L$ and assumed that our variational distribution $q$ factorises in the same way. In the second equality we separated out the $\*z_L$ terms, which can be done because $\*z_L$ does not depend on any of the other $\*z_\ell$'s and thus the remaining integrals in the KL-divergence integrate to one. 

We now define the support of the random variable $\*z_\ell$ as $\mathcal{Z}_\ell$ and proceed to further simplify the right-most KL term of Eq.~\ref{eq:plugged_in_model}. 
\begin{equation}\label{eq:second_kl_simplification}
\begin{split}
& \int_{\mathcal{Z}_L} \dots\int_{\mathcal{Z}_1} q(\*z_1,\dots,\*z_L|\*s,\*x) \log \frac{q(\*z_1,\dots,\*z_{L-1} |\*z_{L},\*s,\*x)}{p(\*z_1,\dots,\*z_{L-1} |\*z_{L},\*x)} d\*z_1 \dots d\*z_L  \\
& = \int_{\mathcal{Z}_L} \dots\int_{\mathcal{Z}_1} q(\*z_1,\dots,\*z_L|\*s,\*x) \log \frac{\prod_{\ell=1}^{L-1} q(\*z_\ell|\*z_{\ell+1},\*s,\*x)}{\prod_{\ell=1}^{L-1} p(\*z_\ell|\*z_{\ell+1},\*x)} d\*z_1 \dots d\*z_L  \\
& = \int_{\mathcal{Z}_L}\dots\int_{\mathcal{Z}_1} q(\*z_1,\dots,\*z_L|\*s,\*x) \left[ \sum_{\ell=1}^{L-1} \log \frac{q(\*z_\ell|\*z_{\ell+1},\*s,\*x)}{p(\*z_\ell|\*z_{\ell+1},\*x)}\right]d\*z_1 \dots d\*z_L \\
& = \sum_{\ell=1}^{L-1} \int_{\mathcal{Z}_L}\dots\int_{\mathcal{Z}_{\ell+1}} q(\*z_{\ell+1},\dots,\*z_{L}|\*s,\*x) d\*z_{\ell+2}\dots d\*z_L \int_{\mathcal{Z}_\ell} q(\*z_\ell|\*z_{\ell+1}) \\ 
& \,\,\,\,\,\, \int_{\mathcal{Z}_{\ell-1}}\dots\int_{\mathcal{Z}_1} q(\*z_1,\dots, \*z_{\ell-1}|\*z_\ell,\*s,\*x) d\*z_{1}\dots d\*z_{\ell-1} \log \frac{q(\*z_\ell|\*z_{\ell+1},\*s,\*x)}{p(\*z_\ell|\*z_{\ell+1},\*x)} d\*z_\ell d\*z_{\ell+1} \\
& = \sum_{\ell=1}^{L-1} \int_{\mathcal{Z}_{\ell+1}} q(\*z_{\ell+1}|\*s,\*x) \int_{\mathcal{Z}_\ell} q(\*z_\ell|\*z_{\ell+1}) \log \frac{q(\*z_\ell|\*z_{\ell+1},\*s,\*x)}{p(\*z_\ell|\*z_{\ell+1},\*x)}d\*z_\ell d\*z_{\ell+1} \\
& = \sum_{\ell=1}^{L-1} \E_{q(\*z_{\ell+1}|\*s,\*x)} \left[ \KL\left[ q(\*z_\ell|\*z_{\ell+1}, \*s, \*x) || p(\*z_\ell|\*z_{\ell+1}, \*x) \right] \right].
\end{split}
\end{equation}
For the first equality above, we used the definition of our graphical model on the hidden variables inside the log term and assumed that $q$ factorises in the same way. For the second equality, we used the log property. To obtain the third equality, we moved the sum out of the integrals and regrouped the integrals based on the index of the sum. Note that the integrals corresponding to the terms $\*z_{\ell-1}$ to $\*z_1$ can be moved inside because the log term does not depend on them. For the fourth equality, we noticed that the integral over $q(\*z_1,\dots,\*z_{\ell-1}|\*z_\ell,\*s,\*x)$ evaluates to one because $q$ must be a normalised probability distribution. We also marginalised out the variables $\*z_{\ell-1}$ to $\*z_{L}$ in the integral over $q(\*z_{\ell+1},\dots,\*z_{L}|\*s,\*x)$. Lastly, we used the definitions of the expectation and the KL-divergence to arrive at the final term. 

By plugging back the simplifications obtained in Eqs.~\ref{eq:plugged_in_model} and \ref{eq:second_kl_simplification} into Eq. \ref{eq:var_decomp} we arrive at the expression for the evidence lower bound given in Eq.~2 in the main article. 

\section{Additional Samples}
\label{sec:appendix_samples}
\setcounter{figure}{0}   

\begin{figure}
\centering
\includegraphics[width=0.99\textwidth]{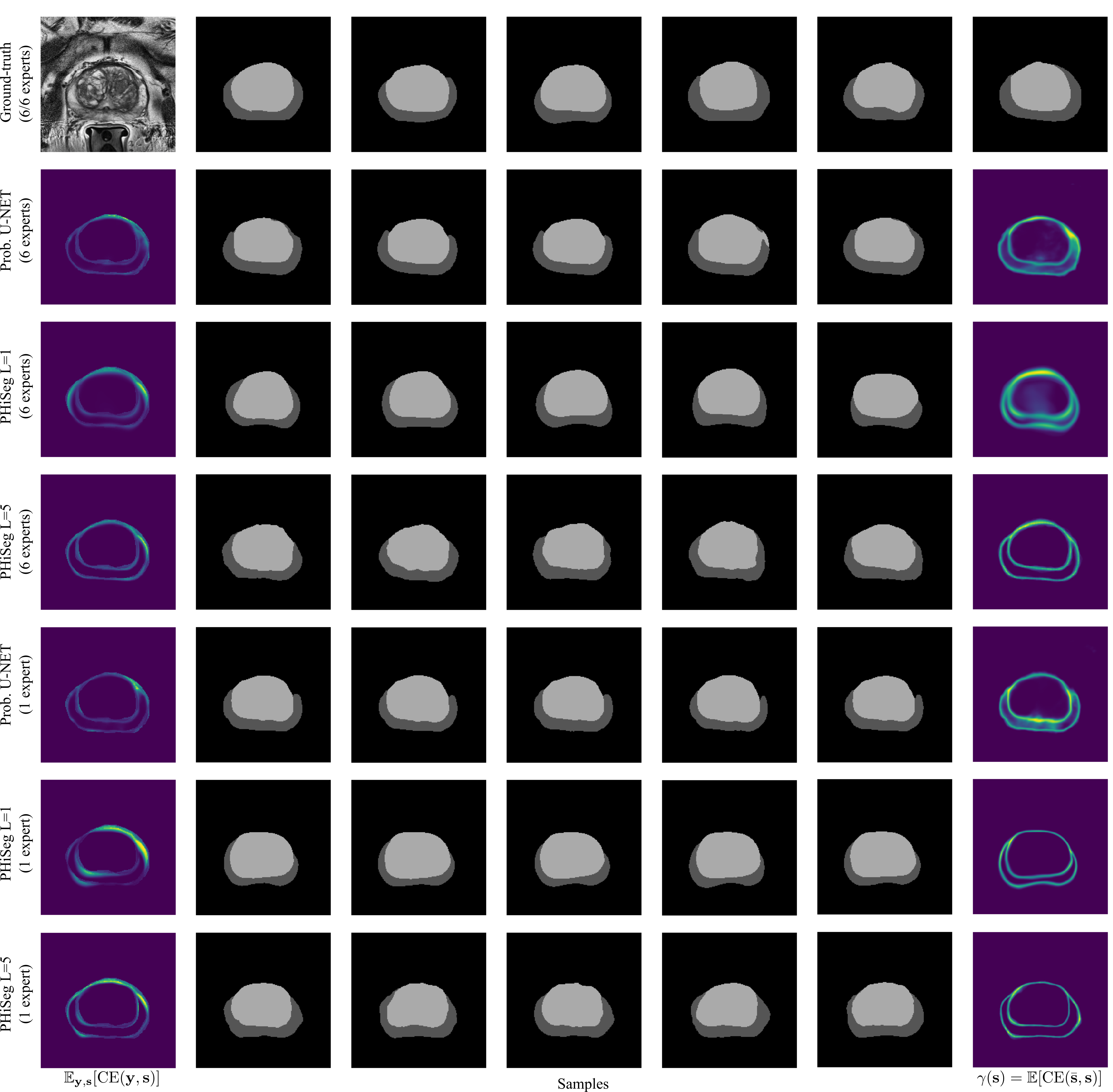}  
\caption{Prostate MR slice with comparatively small inter-expert disagreement. Ground-truth annotations and samples for all evaluated methods trained with masks from 6 or 1 experts(s) are shown. Average error maps $\E_{\*y,\*s}[\CE(\*y, \*s)]$ and $\gamma$-maps ($\E[\CE(\bar{\*s}, \*s)]$) for each model are shown in the left and right-most column, respectively. The first and last column should match in a perfect model.}
\label{fig:sup_prostate}
\end{figure}

\begin{figure}
\centering
\includegraphics[width=0.99\textwidth]{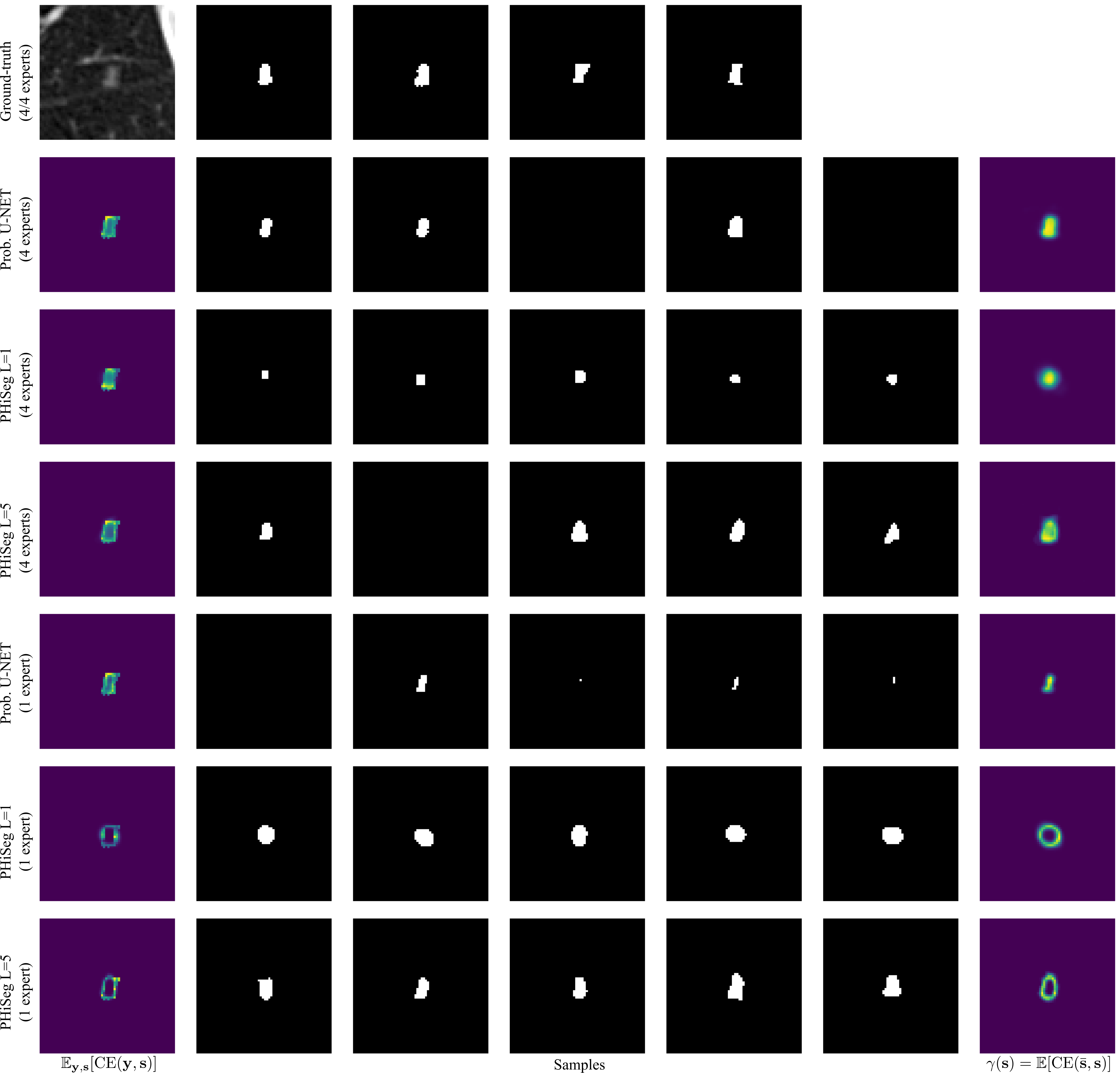}  
\caption{Thoracic lesion in CT image (zoomed in with factor $\times 2$ for better visualisation). Ground-truth annotations and samples for all evaluated methods trained with masks from 4 or 1 experts(s) are shown. Average error maps $\E_{\*y,\*s}[\CE(\*y, \*s)]$ and $\gamma$-maps ($\E[\CE(\bar{\*s}, \*s)]$) for each model are shown in the left and right-most column, respectively. The first and last column should match in a perfect model.}
\label{fig:sup_lidc1} 
\end{figure}

\begin{figure}
\centering
\includegraphics[width=0.99\textwidth]{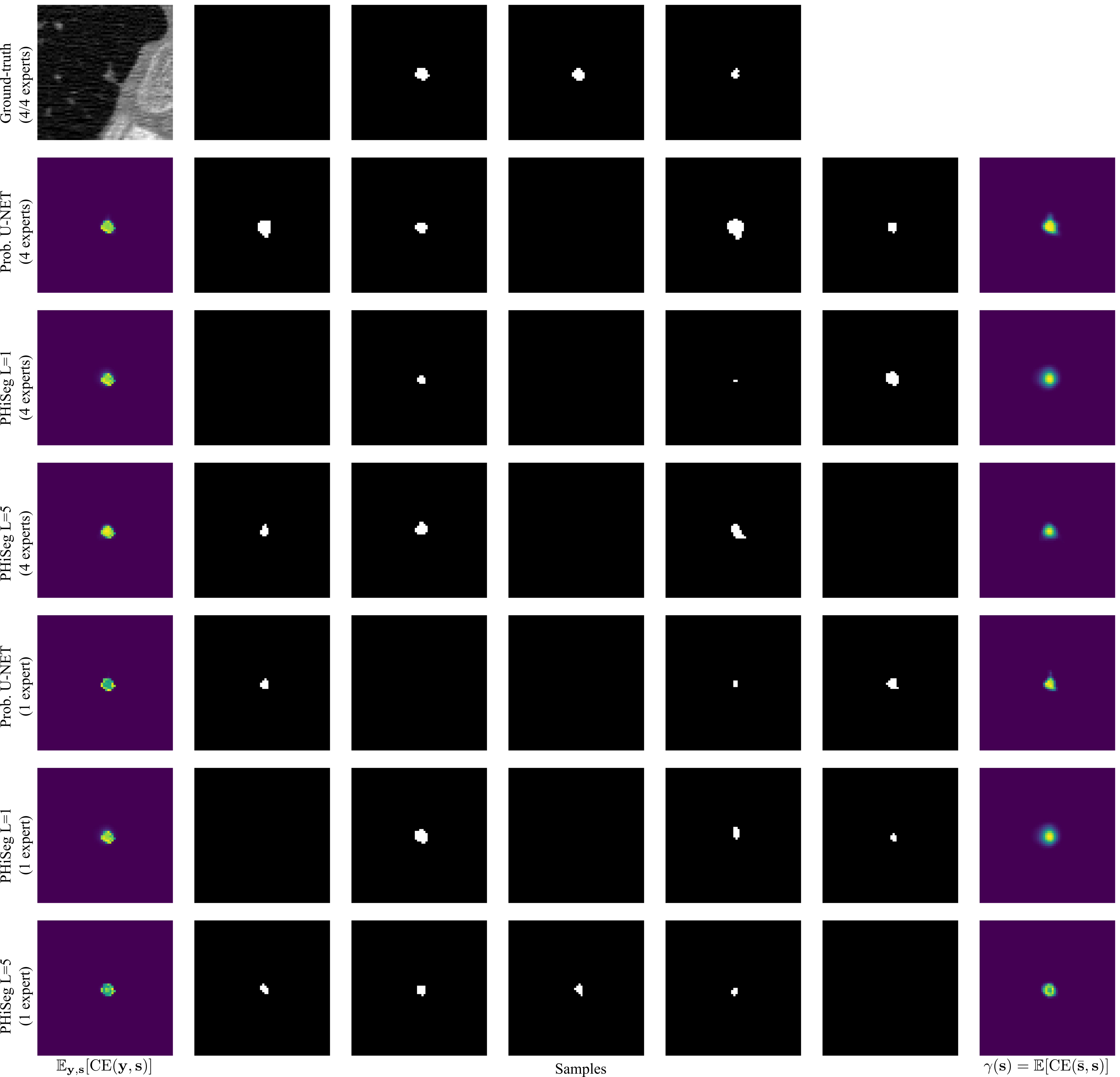}  
\caption{Thoracic lesion in CT image with inter-expert disagreement over presence or absence of lesion (zoomed in with factor $\times 2$ for better visualisation). Ground-truth annotations and samples for all evaluated methods trained with masks from 4 or 1 experts(s) are shown. Average error maps $\E_{\*y,\*s}[\CE(\*y, \*s)]$ and $\gamma$-maps ($\E[\CE(\bar{\*s}, \*s)]$) for each model are shown in the left and right-most column, respectively. The first and last column should match in a perfect model.}
\label{fig:sup_lidc2}
\end{figure}

In this section, we show additional samples generated using all of the investigated methods for the prostate dataset (see Fig.~\ref{fig:sup_prostate}), and the LIDC dataset (see Figs.~\ref{fig:sup_lidc1} and \ref{fig:sup_lidc2}). 

Fig.~\ref{fig:samples_prostate} in the main article shows examples for a prostate with large inter-expert disagreements. In contrast, Fig.~\ref{fig:sup_prostate} is showing an example where the annotation disagreements were relatively smaller. 

Fig.~\ref{fig:sup_lidc1} shows a thoracic lesion with disagreements regarding the shape of the lesion, while Fig.~\ref{fig:sup_lidc2} is showing an example where the disagreement is over presence or absence of the lesion. 

\end{document}